\documentclass[12pt]{article}

\usepackage{amsfonts}
\usepackage{amsmath}
\usepackage{amssymb}
\usepackage{subfigure}
\usepackage{graphicx}
\usepackage{caption}
\usepackage{color}

\usepackage{graphicx,epsf}
\usepackage{enumerate}
\usepackage{natbib}
\usepackage{url} 

\newcommand{\blind}{0}

\newtheorem{prop}{\sc Proposition}[section]

\def \bfit0{\mbox{\textbf{\textit{0}}}}
\def \bfita{\mbox{\textbf{\textit{a}}}}

\def \bfith{\mbox{\textbf{\textit{h}}}}
\def \bfitr{\mbox{\textbf{\textit{r}}}}

\def \bfitu{\mbox{\textbf{\textit{u}}}}
\def \bfitv{\mbox{\textbf{\textit{v}}}}
\def \bfitw{\mbox{\textbf{\textit{w}}}}
\def \bfitx{\mbox{\textbf{\textit{x}}}}

\def \bfitz{\mbox{\textbf{\textit{z}}}}

\def \sbfitu{\mbox{\scriptsize\textbf{\textit{u}}}}

\def \sbfitw{\mbox{\scriptsize\textbf{\textit{w}}}}
\def \sbfitx{\mbox{\scriptsize\textbf{\textit{x}}}}

\def \tbfitx{\mbox{\tiny\textbf{\textit{x}}}}

\def \bfitA{\mbox{\textbf{\textit{A}}}}

\def \bfitU{\mbox{\textbf{\textit{U}}}}
\def \sbfitU{\mbox{\scriptsize\textbf{\textit{U}}}}

\def \bfitW{\mbox{\textbf{\textit{W}}}}
\def \sbfitW{\mbox{\scriptsize\textbf{\textit{W}}}}
\def \bfitX{\mbox{\textbf{\textit{X}}}}
\def \sbfitX{\mbox{\scriptsize\textbf{\textit{X}}}}
\def \bfitZ{\mbox{\textbf{\textit{Z}}}}

\def \bfbeta{\mbox{\boldmath $\beta$}}

\def \bfgamma{\mbox{\boldmath $\gamma$}}
\def \bfmu{\mbox{\boldmath $\mu$}}

\def \bftheta{\mbox{\boldmath $\theta$}}

\def \bfxi{\mbox{\boldmath $\xi$}}

\def \M{\mathbf{M}}

\def \I{\mathbf{I}}
\def \H{\mathbf{H}}

\def \W{\mathbf{W}}

\def \bfGamma{\mbox{\boldmath $\Gamma$}}
\def \bfXi{\mbox{\boldmath $\Xi$}}

\def \bfSigma{\mbox{\boldmath $\Sigma$}}

\begin{document}

\def\spacingset#1{\renewcommand{\baselinestretch}%
{#1}\small\normalsize} \spacingset{1}


\if0\blind
{ \title{\bf Prediction analysis for microbiome sequencing data}
  \author{Tao Wang \\
  \small Department of Bioinformatics and Biostatistics, Shanghai Jiao Tong University\\
  Can Yang \\
  \small Department of Mathematics, Hong Kong Baptist University\\
  and \\
  Hongyu Zhao
    \\
    \small Department of Biostatistics, Yale University}
  \maketitle
} \fi

\if1\blind
{
  \bigskip
  \bigskip
  \bigskip
  \begin{center}
    {\LARGE\bf Title}
\end{center}
  \medskip
} \fi

\bigskip
\begin{abstract}
One primary goal of human microbiome studies is to predict host traits based on human microbiota. However, microbial community sequencing data present significant challenges to the development of statistical methods. In particular, the samples have different library sizes, the data contain many zeros and are often over-dispersed. To address these challenges, we introduce a new statistical framework, called predictive analysis in metagenomics via inverse regression (PAMIR). An inverse regression model is developed for over-dispersed microbiota counts given the trait, and then a prediction rule is constructed by taking advantage of the dimension-reduction structure in the model. An efficient Monte Carlo expectation-maximization algorithm is designed for carrying out maximum likelihood estimation. We demonstrate the advantages of PAMIR through simulations and a real data example.
\end{abstract}

\noindent%

\vfill

\newpage
\spacingset{1.45} 

\section{Introduction}

Next generation sequencing technologies have allowed high-throughput surveys of human-associated microbial communities \citep{turnbaugh2007human}. In these surveys, one of the most important research efforts  is to predict host traits, such as disease states, based on human microbiota \citep{knights2011human}. Recent studies have shown that many microbes are either harmless or of benefit to the host, indicating the potential of microbiota-based characterization of host phenotypes \citep{cho2012human}.

Existing statistical methods for understanding the relationship between a trait of interest and the human microbiota can be roughly classified into two major categories: those that are test-based and those that are model-based. In the first category, most methods test on single microbes one at a time, followed by a proper adjustment for multiple testing; see, for example, \citet{le2013richness}. However, these methods discount the inherent properties of microbiota data \citep{li2015microbiome}. Furthermore, multiple testing can result in a loss of power when associations are weak. To deal with these issues, distance-based methods evaluate the association of the overall microbiota composition with the trait \citep{charlson2010disordered}. By partitioning the distance matrix among sources of variation, the statistical significance can be tested by permutation \citep{mcardle2001fitting}. 
In the second category, regression models are used to decipher the relationship between the trait and the microbiota. For example, \citet{lin2014variable} adopted the linear log-contrast model for compositional data, and
developed a penalized method for removing unimportant microbes. \citet{garcia2014identification} grouped the microbiota from phylum to species level, and proposed a new variable selection method to identify important features at multiple taxonomic levels. Taking into account the phylogenetic relationships among the microbes, \citet{tanaseichuk2014phylogeny} proposed a novel method for classifying microbial communities. Recently, \citet{zhao2015testing} introduced a regression-based test for assessing the association between the trait and the microbial diversity. Although tests are simple and powerful, methods under a principled regression framework deal explicitly with estimation and prediction, are interpretable, and are easily extended in many ways.

The nature of microbiota data creates significant challenges and great opportunities for the development of statistical methods. In particular, the samples have different sequencing depths, the data contain many zeros and are often over-dispersed \citep{weiss2015effects}. Some progress has been made to address these challenges. To account for library size differences, the common approach is to use simple proportions. This makes sense, since the observed sequences are relative abundances. Because the proportions must sum to one, the data are compositional. Elegant statistical theory is available for analyzing compositional data, and inferences based on log-ratios are popular in current practice; see, for example, \citet{friedman2012inferring} and \citet{lin2014variable}. A drawback of these methods is that taking logarithms of the proportions is problematic in the presence of zeros. One can add a pseudo positive constant to the raw counts, but the choice of the constant is often arbitrary. Recent publications have advocated modeling multivariate count data directly. The classical multinomial-logit regression model is commonly used, but its mean-variance structure is very restrictive. To allow for over-dispersion, variants of the multinomial distribution, such as the Dirichlet-multinomial distribution and the additive logistic normal multinomial distribution, have been used as probability models \citep{chen2013variable, xia2013logistic}. These models can handle zero counts automatically, and statistical inferences are made conditional on the total count determined by the sampling depth. See \citet{li2015microbiome} for a review of data characteristics of metagenomic studies as well as computational and statistical challenges. 

The goal of this paper is to build flexible statistical models for predicting a host phenotype based on the human microbiota. Although there is a vast literature on predictive models within statistics and machine learning, thus far, few studies with successful microbiota-based prediction of outcomes have been reported \citep{gevers2014treatment, teng2015prediction}. One explanation is that the aforementioned properties of microbial community data make that goal difficult to obtain using traditional approaches. By employing a technique called inverse regression \citep{li1991sliced}, we present a new statistical methodology, called predictive analysis in metagenomics via inverse regression (PAMIR). PAMIR reverses the roles of the trait and the microbiota. Instead of regressing the trait on the microbiota, it performs a regression of the microbiota on the trait. In Section \ref{sec_mod}, we introduce an inverse regression model for over-dispersed microbiota counts given the trait. In Section \ref{sec_pred}, we take advantage of the dimension-reduction structure in the model to construct a prediction rule at the population level. Estimation of model parameters is considered in Section \ref{sec_est}, where we develop a Monte Carlo expectation-maximization algorithm. Some simulations are reported in Section \ref{sec_simu}. In Section \ref{sec_data}, we apply the proposed methods to a gut microbiome data set. Finally, a discussion is given in Section \ref{sec_disc}.

\section{An inverse regression model}\label{sec_mod}

Throughout this paper, random variables are denoted by uppercase letters, and their values are written in lowercase. Let $\bfitX_y = (X_{y1}, \ldots, X_{yp})^{\top}$ denote a random vector distributed as $\bfitX \mid (Y = y)$. Suppose that $\bfitx_y = (x_{y1}, \ldots, x_{yp})^{\top}$ is a draw from $\bfitX_y$. Let $m_y = \sum_{j = 1}^p x_{yj}$. The development of this paper is conditional on $m_y$. Define
$$\binom{m_y}{\bfitx_y} = \frac{\Gamma(m_y + 1)}{\prod_{j = 1}^p \Gamma(x_{yj} + 1)},$$
where $\Gamma(\cdot)$ is the gamma function. A popular multivariate model for $\bfitX_y$ has probability mass function
\begin{eqnarray}\label{multinomial}
\binom{m_y}{\bfitx_y} \prod_{j = 1}^p z_{yj}^{x_{yj}},
\end{eqnarray}
where $\bfitz_y = (z_{y1}, \ldots, z_{yp})^{\top} \in \mathbb{R}^p$ is a vector of probabilities such that $\sum_{j = 1}^p z_{yj} = 1$.

It is easy to see that in (\ref{multinomial}) no direct use is made of the response, which plays the role of an implicit conditioning argument. This is clearly a drawback, since we are interested in understanding the nature of the relationship between the covariates and the response. One way to deal with this issue is by modeling $\bfitz_y$ \citep{cook2007fisher, taddy2010multinomial}. Specifically, we assume that
\begin{equation}\label{prob}
z_{yj} = \frac{\exp(a_j + \bfgamma_j^{\top} \bfbeta \bfith_y)}{\sum_{k = 1}^{p}\exp(a_k + \bfgamma_k^{\top} \bfbeta \bfith_y)},
\end{equation}
where $a_j \in \mathbb{R}, \bfgamma_j \in \mathbb{R}^{d}, \bfbeta \in \mathbb{R}^{d \times r}$ has rank $d \leq \min(p, r)$, and $\bfith_y  \in \mathbb{R}^r$ is a known vector-valued function of $y$. Usually, we require that $\bfith_y$ contain a reasonably flexible set of basis functions. By convention, we set $a_p = 0$ and $\bfgamma_p$ to be the $d$-vector of zeros.

Another problem with (\ref{multinomial}) is its difficulty in modeling over-dispersion, which is a well-known feature of count data in microbiome studies. To account for over-dispersion, the standard convention is to assume that the vector of proportions $\bfitz_{y}$ is itself random with some distribution \citep{chen2013variable, xia2013logistic}. Under (\ref{prob}), we can achieve this by assuming that $\bfita = (a_1, \ldots, a_{p - 1})^{\top} \in \mathbb{R}^{p - 1}$ is a realization of $\bfitA = (A_1, \ldots, A_{p - 1})^{\top}$.

Let $W_{yj} = A_j + \bfgamma_j^{\top} \bfbeta \bfith_y$ and $\bfitW_{y} = (W_{y1}, \ldots, W_{y(p - 1)})^{\top}$. We assume that $\bfitA$ is normally distributed with mean vector $\bfmu$ and covariance matrix $\bfSigma$, and is independent of $Y$. Then we can write
\begin{equation}\label{linear}
\bfitW_y = \bfmu + \bfGamma \bfbeta \bfith_y + \bfxi,
\end{equation}
where $\bfGamma = (\bfgamma_1, \ldots, \bfgamma_{p - 1})^{\top} \in \mathbb{R}^{(p - 1) \times d}$ and $\bfxi = \bfitA - \bfmu$. For a positive integer $k$, denote by $\I_{k}$ the $k \times k$ identity matrix. Without loss of generality, we assume that $\bfGamma^{\top} \bfSigma^{-1} \bfGamma = \I_d$. In this paper, the subscript $y$ is either used to emphasize the conditional nature of the model, or used to index observations in place of the traditional notation.

\section{Dimension reduction and prediction}\label{sec_pred}

Before we continue, we need a definition. Let $\mathbb{S}^{p - 1}$ denote the $(p - 1)$-dimensional simplex. We define the transformation of $\bfitz = (z_1, \ldots, z_p) \in \mathbb{S}^{p - 1}$ to $\mathbb{R}^{p - 1}$ as
\begin{equation*}
\phi(\bfitz) = \left\{\log\left(\frac{z_{1}}{z_{p}}\right), \ldots, \log\left(\frac{z_{p-1}}{z_{p}}\right)\right\}.
\end{equation*}
This transformation is a bijection, and is called the additive log-ratio transformation \citep{aitchison1986statistical}. It can be shown that
\begin{equation}\label{alr}
\phi(\bfitZ_y) = \bfitW_y,
\end{equation}
and hence
\begin{equation}\label{alt}
\bfitZ_y = \phi^{-1}(\bfitW_y),
\end{equation}
where $\phi^{-1}$ denotes the inverse transformation of $\phi$.

We have the following proposition, the proof of which can be found in the Appendix. 
\begin{prop}\label{prop}
Under (\ref{linear}), $Y$ is independent of $\bfitW$ given $\bfGamma^{\top} \bfSigma^{-1} \bfitW$.
\end{prop}

According to this proposition, $\bfitW$ can be replaced by $\bfGamma^{\top} \bfSigma^{-1} \bfitW$, without loss of information on the regression of $Y$ on $\bfitW$. The latter is called a sufficient reduction in the dimension-reduction literature \citep{cook1998regression}. However, unlike in the standard framework of dimension reduction, $\bfitW$ is unobservable here.

\citet{taddy2010multinomial} proposed multinomial inverse regression for text analysis. His method was based on conditional sufficiency. Specifically, Proposition 3.2 of \citet{taddy2010multinomial} stated that, given $\bfxi$ and $\bfGamma^{\top} \bfitX$, $Y$ is independent of $\bfitX$. However, the sufficiency of $\bfGamma^{\top} \bfitX$ can not be justified without a conditioning argument, making subsequent forward regression unreliable. In our framework, we treat $\bfxi$ unconditionally. This has an important implication for microbiota data. Conditioning on $Y=y$, the proportions follow Aitchison's logistic normal distribution for compositional data \citep{aitchison1986statistical}. Another difference between Taddy's method and ours is that, while the coordinate vectors were pre-specified in his model, they are modeled parametrically via $\bfbeta \bfith_y$ in (\ref{linear}). 

To predict a future observation of $Y$ associated with a new observed vector of $\bfitX$, we use the forward regression mean function $E(Y \mid \bfitX)$. 
From (\ref{multinomial}) and (\ref{alt}) we have
\begin{eqnarray*}
E(Y \mid \bfitX) = E\{E(Y \mid \bfitW) \mid \bfitX\}.
\end{eqnarray*}
In this paper we construct a prediction rule by taking advantage of this observation. Loosely speaking, our method contains two parts: estimation of $E(Y \mid \bfitX)$ when $E(Y \mid \bfitW)$ is known, and estimation of $E(Y \mid \bfitW)$. In either part, we rely on estimates of $\bfGamma$ and other parameters. For the moment we assume that these parameters are known. We defer the estimation problem to Section \ref{sec_est}.

A computational approach of estimating $E(Y \mid \bfitX = \bfitx)$, when $E(Y \mid \bfitW)$ is known, is to draw samples from the conditional distribution of $E(Y \mid \bfitW)$ given $\bfitX = \bfitx$, and then use the sample mean as the predicted value. By (\ref{multinomial}) and (\ref{alt}), the conditional density of $\bfitW$ given $\bfitX$ is
\begin{eqnarray}
\nonumber
f_{\sbfitW \mid \sbfitX}(\bfitw \mid \bfitx)
&\propto& f_{\sbfitX \mid \sbfitW}(\bfitx \mid \bfitw) \times f_{\sbfitW}(\bfitw) \\ \nonumber
&=& \frac{\prod_{j = 1}^{p - 1} \exp(x_{j} w_{j})}{\{\prod_{j = 1}^{p - 1} \exp(w_{j}) + 1\}^{m}} \times f_{\sbfitW}(\bfitw) .
\end{eqnarray}
Here $m = \sum_{j = 1}^p x_j$. We can approximate $f_{\sbfitW}(\bfitw)$ by
\begin{eqnarray*}
\frac{1}{n}\sum_y f_{\sbfitW_y}(\bfitw_y),
\end{eqnarray*}
where the summation is over the observed values of $Y$. Hence, $f_{\sbfitW \mid \sbfitX}(\bfitw \mid \bfitx)$ is approximately proportional to
$$\frac{\prod_{j = 1}^{p - 1} \exp(x_{j} w_{j})}{\{\prod_{j = 1}^{p - 1} \exp(w_{j}) + 1\}^{m}} \times \left\{\sum_y f_{\sbfitW_y}(\bfitw_y)\right\}.$$
This allows us to draw samples from $f_{\sbfitW \mid \sbfitX}(\bfitw \mid \bfitx)$, denoted by $\mathcal{W}_{\sbfitx}$, using, for example, the Metropolis--Hastings algorithm, and then estimate $E(Y \mid \bfitX)$ by
\begin{eqnarray}\label{pred_11}
\tilde{E}(Y \mid \bfitX) = \frac{1}{|\mathcal{W}_{\sbfitx}|} \sum_{\sbfitw \in \mathcal{W}_{\tbfitx}} E(Y \mid \bfitW  = \bfitw),
\end{eqnarray}
where $|\mathcal{W}_{\sbfitx}|$ denotes the size of $\mathcal{W}_{\sbfitx}$.

Let $\bfitU = \bfGamma^{\top}\bfSigma^{-1} \bfitW$. We have $E(Y \mid \bfitW) = E(Y \mid \bfitU)$. When the dimension of $\bfitU$ is low, there are a variety of efficient nonparametric methods for estimating $E(Y \mid \bfitU)$; see, for example, \citet{hardle1990applied}. Under (\ref{linear}), a simple alternative can be constructed by noting that
\begin{eqnarray*}
E(Y \mid \bfitU = \bfitu) 
= \frac{\int  y f_{\sbfitU \mid Y}(\bfitu \mid y) f(y) \operatorname{d} y}{f_{\sbfitU}(\bfitu)}
= \frac{E\{Y f_{\sbfitU \mid Y}(\bfitu \mid Y)\}}{E\{f_{\sbfitU \mid Y}(\bfitu \mid Y)\}}.
\end{eqnarray*}
An estimate is then obtained by replacing the expectations by averages over the observed data:
\begin{eqnarray}\label{pred_12}
\tilde{E}(Y \mid \bfitU = \bfitu) = \frac{\sum_{y} y f_{\sbfitU \mid Y}(\bfitu \mid y)}{\sum_y f_{\sbfitU \mid Y}(\bfitu \mid y)}.
\end{eqnarray}
Combining (\ref{pred_11}) and (\ref{pred_12}), the predicted value of $Y$ at the given value $\bfitx$ of $\bfitX$ is given by
\begin{eqnarray}\label{pred_1}
\hat{E}(Y \mid \bfitX = \bfitx) = \frac{1}{|\mathcal{U}_{\sbfitx}|}\sum_{\sbfitu \in \mathcal{U}_{\tbfitx}} \frac{\sum_{y} y f_{\sbfitU \mid Y}(\bfitu \mid y)}{\sum_y f_{\sbfitU \mid Y}(\bfitu \mid y)},
\end{eqnarray}
where $\mathcal{U}_{\sbfitx} = \{\bfGamma^{\top}\bfSigma^{-1}\bfitw, \bfitw \in \mathcal{W}_{\sbfitx}\}$.


\section{Parameter estimation}\label{sec_est}

Let $\bftheta = \{\bfmu, \bfGamma, \bfbeta, \bfSigma\}$. We wish to estimate $\bftheta$, based on a random sample of size $n$ from the joint distribution of $Y$ and $\bfitX$. Since no closed-form likelihood function is available, it is usually not possible to find the closed-form maximum likelihood estimate of $\bftheta$. In this section, we propose an Expectation-Maximization (EM) algorithm for finding locally maximum likelihood estimates.

By (\ref{alt}), the complete data log-likelihood can be written as
\begin{eqnarray}
\nonumber
l(\bftheta)
&=& \log\left[\prod_{y} \left\{f_{\sbfitX_y \mid \sbfitW_y}(\bfitx_y \mid \bfitw_y) f_{\sbfitW_y}(\bfitw_y; \bftheta)\right\}\right] \\
&=& \sum_{y} \log\{f_{\sbfitX_y \mid \sbfitW_y}(\bfitx_y \mid \bfitw_y)\} + \sum_{y} \log\{f_{\sbfitW_y}(\bfitw_y; \bftheta)\},
\end{eqnarray}
where the product or sum is over the observed values of $Y$.

The EM algorithm seeks to find the maximum likelihood estimate of $\bftheta$ by iteratively applying an Expectation (E) step and a Maximization (M) step. Let
$$\bftheta^{(t - 1)} = \{\bfmu^{(t - 1)}, \bfGamma^{(t - 1)}, \bfbeta^{(t - 1)}, \bfSigma^{(t - 1)}\}$$
be the parameters at the $(t - 1)$th M step. In the $t$th E step, we calculate the conditional expectation of $l(\bftheta)$ with respect to the distribution of $\bfitW_y \mid \bfitX_y$ governed by $\bftheta^{(t - 1)}$:
\begin{eqnarray}
\nonumber
Q(\bftheta; \bftheta^{(t - 1)}) &=& E\{l(\bftheta)\} = c + E\left[\sum_{y} \log\{f_{\sbfitW_y}(\bfitw_y; \bftheta)\}\right].
\end{eqnarray}
Here, $c$ is an irrelevant constant.

By (\ref{linear}) and the normality of $\bfxi$,
\begin{eqnarray}
\nonumber
&&\log\{f_{\sbfitW_y}(\bfitw_y; \bftheta)\} \\ \nonumber
&=& -\frac{p - 1}{2}\log(2\pi) - \frac{1}{2}\log\{\operatorname{det}(\bfSigma)\} - \frac{1}{2} (\bfitw_y - \bfmu - \bfGamma\bfbeta \bfith_y)^{\top} \bfSigma^{-1} (\bfitw_y - \bfmu - \bfGamma\bfbeta \bfith_y).
\end{eqnarray}
Hence
\begin{eqnarray}
\nonumber
 Q(\bftheta; \bftheta^{(t - 1)}) &=& c - \frac{n(p - 1)}{2}\log(2\pi) - \frac{n}{2}\log\{\operatorname{det}(\bfSigma)\} \\
&& \quad -\frac{1}{2}\sum_{y} E\{(\bfitw_y - \bfmu - \bfGamma\bfbeta \bfith_y)^{\top} \bfSigma^{-1} (\bfitw_y - \bfmu - \bfGamma\bfbeta \bfith_y)\}.
\end{eqnarray}
To compute the conditional expectations, we use the Metropolis--Hastings (MH) algorithm. Note that
\begin{eqnarray}
\nonumber
&& f_{\sbfitW_y \mid \sbfitX_y}(\bfitw_y \mid \bfitx_y; \bftheta^{(t - 1)}) \\ \nonumber
&\propto& f_{\sbfitX_y \mid \sbfitW_y}(\bfitx_y \mid \bfitw_y) \times f_{\sbfitW_y}(\bfitw_y; \bftheta^{(t - 1)}) \\ \nonumber
&\propto& \frac{\prod_{j = 1}^{p - 1} \exp(x_{yj} w_{yj})}{\{\prod_{j = 1}^{p - 1} \exp(w_{yj}) + 1\}^{m_y}} \times \exp\left[-\frac{1}{2} \bfitr_y^{(t - 1)\top} \{\bfSigma^{(t - 1)}\}^{-1} \bfitr_y^{(t - 1)}\right],
\end{eqnarray}
where $\bfitr_y^{(t - 1)} = \bfitw_y - \bfmu^{(t - 1)} - \bfGamma^{(t - 1)\top} \bfbeta^{(t - 1)} \bfith_y$. We sample from this conditional distribution as follows. In the $r$th MH step, we draw a candidate vector $\bfitw_y^*$ from a multivariate normal distribution with mean vector $\bfitw_y^{(r - 1)}$ and covariance matrix $\I_{p - 1}$, and compute the acceptance ratio
$$\kappa = \min\left\{1, \frac{f_{\sbfitW_y \mid \sbfitX_y}(\bfitw_y^* \mid \bfitx_y, \bftheta^{(t - 1)})}{f_{\sbfitW_y \mid \sbfitX_y}(\bfitw_y^{(r - 1)} \mid \bfitx_y, \bftheta^{(t - 1)})}\right\}.$$
We then simulate a random number $u$ from the uniform distribution on the interval $[0, 1]$, and accept the candidate by setting $\bfitw_y^{(r)} = \bfitw_y^*$, if $\kappa \geq u$, and keep the previous value, otherwise. After a burn-in process in which an initial number of samples are thrown away, we use the next $B$ samples, denoted by $\{\bfitw_{y}^{1}, \ldots, \bfitw_{y}^{B}\}$, to calculate the conditional expectation in the E step. Ignoring constants, this leads to the quantity
\begin{eqnarray}
\nonumber
&& \tilde{Q}(\bftheta; \bftheta^{(t - 1)}) \\ \nonumber
&=& - \frac{n}{2}\log\{\operatorname{det}(\bfSigma)\} - \frac{1}{2B}\sum_{y} \sum_{b = 1}^{B}\{(\bfitw_y^b - \bfmu - \bfGamma\bfbeta \bfith_y)^{\top} \bfSigma^{-1} (\bfitw_y^b - \bfmu - \bfGamma\bfbeta \bfith_y)\}.
\end{eqnarray}

In the $t$th M step, we maximize $\tilde{Q}(\bftheta; \bftheta^{(t - 1)})$ over $\bftheta$. Without loss of generality, assume that $\{\bfith_y\}$ are centered, that is, $\sum_{y} \bfith_y$ is the $r$-vector of zeros. Let $\bar{\bfitw}_y = B^{-1}\sum_{b = 1}^B\bfitw_y^b$ and $\bar{\bfitw} = n^{-1} \sum_{y} \bar{\bfitw}_y$. For fixed $(\bfGamma, \bfbeta, \bfSigma)$, $\tilde{Q}(\bftheta; \bftheta^{(t - 1)})$ is maximized over $\bfmu$ by
$$\bfmu^{(t)} = \bar{\bfitw}.$$
Substituting $\bfmu^{(t)}$ into $\tilde{Q}$, we obtain
\begin{eqnarray}
\nonumber
&&\tilde{Q}(\{\bfmu^{(t)}, \bfGamma, {\bfbeta}, \bfSigma\}; \bftheta^{(t - 1)}) \\ \nonumber
&=& - \frac{n}{2}\log\{\operatorname{det}(\bfSigma)\} - \frac{1}{2B}\sum_{y} \sum_{b = 1}^{B}\{(\bfitw_{y}^{b} - \bar{\bfitw} - \bfGamma{\bfbeta} \bfith_y)^{\top} \bfSigma^{-1} (\bfitw_{y}^{b} - \bar{\bfitw} - \bfGamma {\bfbeta} \bfith_y)\}.
\end{eqnarray}
To update $(\bfGamma, \bfbeta, \bfSigma)$, we use an alternating algorithm: we first fix $\bfSigma$ and solve for $(\bfGamma, \bfbeta)$, then we fix $(\bfGamma, \bfbeta)$ and solve for $\bfSigma$, and we iterate between these two steps until the algorithm converges. Let $\bar{\W} = (\bar{\bfitw}_y) \in \mathbb{R}^{(p - 1) \times n}$, $\H = (\bfith_y) \in \mathbb{R}^{r \times n}$, and
\begin{eqnarray*}
\M = (\bar{\W} - \bar{\bfitw} \otimes {\bf 1}_n^{\top}) \H^{\top} (\H\H^{\top})^{-1} \H (\bar{\W} - \bar{\bfitw} \otimes {\bf 1}_n^{\top})^{\top}.
\end{eqnarray*}
where ${\bf 1}_n$ is the $n$-vector of ones. Given $\bfSigma$, the solution for $(\bfGamma, \bfbeta)$ is
\begin{eqnarray}\label{update_gamma}
\tilde{\bfGamma} = \bfSigma^{1/2} \{\bfitv_1(\bfSigma), \ldots, \bfitv_d(\bfSigma)\}
\end{eqnarray}
and
\begin{eqnarray}\label{update_beta}
\tilde{\bfbeta} 
= \{\bfitv_1(\bfSigma), \ldots, \bfitv_d(\bfSigma)\}^{\top} \bfSigma^{-1/2} (\bar{\W} - \bar{\bfitw} \otimes {\bf 1}_n^{\top}) \H^{\top} (\H\H^{\top})^{-1},
\end{eqnarray}
where $\bfitv_j(\bfSigma)$ denotes the $j$th largest eigenvector of $\bfSigma^{-1/2} \M \bfSigma^{-1/2}$ (see the Appendix for details). Given $(\bfGamma, \bfbeta)$, the solution for $\bfSigma$ is
 \begin{eqnarray*}
\tilde{\bfSigma} = \frac{1}{n B}\sum_{y} \sum_{b = 1}^{B}(\bfitw_{y}^{b} - \bar{\bfitw} - \bfGamma\bfbeta \bfith_y) (\bfitw_{y}^{b} - \bar{\bfitw} - \bfGamma\bfbeta \bfith_y)^{\top}.
\end{eqnarray*}

Denote by $\hat{\bftheta}$ the estimate of ${\bftheta}$. In the previous section, we show how to predict $Y$ based on $\bfitX$. The procedure also applies with $\bftheta$ replaced by $\hat{\bftheta}$. Suppose that $\bfitx^*$ is a new observation on $\bfitX$, and $\hat{\mathcal{W}}_{\sbfitx^*}$ is a sample from the conditional distribution of $\bfitW \mid (\bfitX = \bfitx^*)$ indexed by $\hat{\bftheta}$. By (\ref{pred_1}), the predicted value is
\begin{eqnarray}\label{predwithx}
\hat{y}^* = \frac{1}{|\hat{\mathcal{U}}_{\sbfitx^*}|}\sum_{\sbfitu \in \hat{\mathcal{U}}_{\tbfitx^*}} \frac{\sum_{y} y \hat{f}_{\sbfitU \mid Y}(\bfitu \mid y)}{\sum_y \hat{f}_{\sbfitU \mid Y}(\bfitu \mid y)},
\end{eqnarray}
where $\hat{\mathcal{U}}_{\sbfitx^*} = \{\hat{\bfGamma}^{\top}\hat{\bfSigma}^{-1}\bfitw, \bfitw \in \hat{\mathcal{W}}_{\sbfitx^*}\}$, and
$$\hat{f}_{\sbfitU \mid Y}(\bfitu \mid y) \propto \exp\left\{-\frac{1}{2}(\bfitu - \hat{\bfGamma}^{\top} \hat{\bfSigma}^{-1} \hat{\bfmu} - \hat{\bfbeta} \bfith_y)^{\top} (\bfitu - \hat{\bfGamma}^{\top} \hat{\bfSigma}^{-1} \hat{\bfmu} - \hat{\bfbeta} \bfith_y)\right\}.$$

\section{Simulations}\label{sec_simu}

In this section, we conduct a simulation study to examine the behavior of our proposed method, PAMIR. We first generated $Y$ from a standard normal distribution. Given $Y = y$, we then generated $\bfitX_y$ according to (\ref{multinomial}), (\ref{prob}), and a simple version of (\ref{linear}) with $d = 1$:
$$\bfitW_y = \bfGamma v_y + \bfxi,$$
where $v_y$ is a function of $y$. We set $n \in \{50, 100\}$, $p \in \{5, 10, 20\}$, $\bfGamma = (1, 1, -1, -1, 0, \ldots, 0)^{\top} / 2$, $v_y = 10 y$, and $\bfSigma = \I_{p - 1}$. For each data set simulated in this way, we used the procedure in Section \ref{sec_est} to fit the model, with $r = 3$ and $\bfith_y = (y, y^2, y^3)^{\top}$. To evaluate the estimation accuracy, we computed the Euclidean distance, $\|\hat{\bfGamma} - \bfGamma\|_2$, between $\hat{\bfGamma}$ and $\bfGamma$. Furthermore, to assess the performance of the rule (\ref{predwithx}), we calculated the mean squared prediction error at 50 new data points, $\{(\bfitx_k^*, y_k^*), k = 1, \ldots, 50\}$, from the same model:
\begin{eqnarray}\label{PErr}
\operatorname{PErr} = \frac{1}{50}\sum_{k = 1}^{50} (\hat{y}_k^* - y^*_k)^2.
\end{eqnarray}
The results based on 100 data replications are shown in Table \ref{tab1}. PAMIR works well in terms of both estimation and prediction. As expected, the performance gets worse as the sample size decreases, and as the number of covariates becomes larger.

\begin{figure}
 \centerline{\includegraphics[height=3in,width=3in,angle=0]{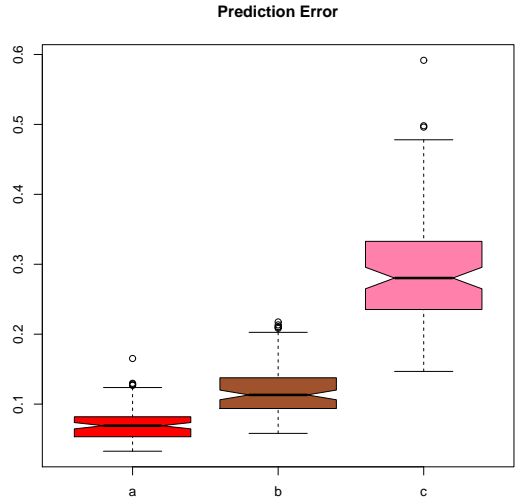}}
\caption{The effect of the mis-specification of $\bfith_y$. (a) $v_y = 10y$; (b) $v_y = 10y + 5|y|$; and (c) $v_y = 10y + 10|y|$.
} \label{fig1}
\end{figure}

We further consider the effect of the choice of $\bfith_y$. The basic simulation scenario was the same, except that in this case we set $v_y = 10(y + c |y|)$ with $c \in \{0.5, 1\}$. Again, each data set was fitted with $\bfith_y = (y, y^2, y^3)^{\top}$, which is now incorrectly specified. The prediction errors are shown in Figure \ref{fig1}. It is clear that our method is not robust to mis-specification of $\bfith_y$. Nevertheless, the general trend suggests that its performance does not deteriorate much if the error of approximating $v_y$ by functions in $\bfith_y$ is small.

\begin{table}[htb!]\caption{Finite sample performance. Reported are the average out of 100 data replications, with standard deviation in parentheses, of the Euclidean distance between $\hat{\bfGamma}$ and $\bfGamma$, and the mean squared prediction error (\ref{PErr}).} \label{tab1} \vspace{-0.3cm}
\centering
 {\small\scriptsize\hspace{12.5cm}
\renewcommand{\arraystretch}{1.2} 
\begin{tabular}{cccc}
\hline
             &           & $\|\hat{\bfGamma} - \bfGamma\|_2$            & PErr           \\
             & $p = 5$   & 0.170 (0.077)   & 0.070  (0.026) \\
   $n = 50$  & $p = 10$  & 0.257 (0.079)   & 0.079  (0.025) \\
             & $p = 20$  & 0.393 (0.076)   & 0.108  (0.029) \\
             & $p = 5$   & 0.121 (0.056)   & 0.059  (0.018) \\
   $n = 100$ & $p = 10$  & 0.199 (0.051)   & 0.072  (0.023) \\
             & $p = 20$  & 0.280 (0.054)   & 0.079  (0.017) \\
\hline
\end{tabular} }
\end{table}

\section{Enterotype data}\label{sec_data}

Clustering of the human gut microbiome into different types, or ``enterotypes", facilitates our understanding of microbial variation in health and disease. Using 22 European samples, 9 Japanese samples, and 2 American samples,
the MetaHIT consortium identified three enterotypes based on the genus compositions of Sanger metagenomes from these samples \citep{arumugam2011enterotypes}. These enterotypes were mostly driven by microbial composition of Bacteroides, Prevotella, and Ruminococcus. Based on multiple-testing corrected correlation analysis, the authors claimed that the enterotypes were not nation or continent specific. We revisited this problem from the viewpoint of classification.

\begin{figure}
 \centerline{\includegraphics[height=3in,width=3in,angle=0]{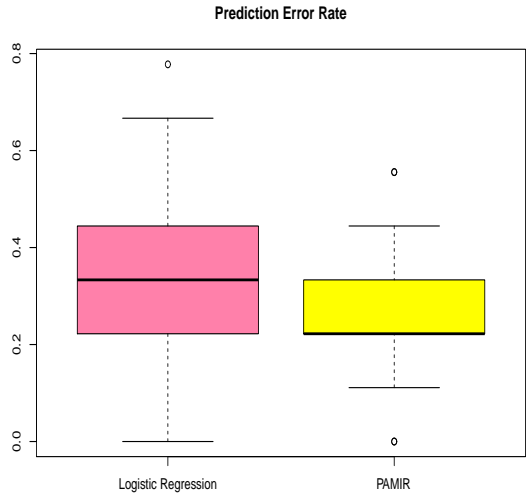}}
\caption{Boxplots of the prediction error rate, over 100 random splits of 33 samples, for logistic regression and PAMIR.
} \label{fig4}
\end{figure}

Specifically, we labeled 22 European individuals as class 1 and the 11 non-Europeans as class 0. We then used the rule (\ref{predwithx}) to predict the class label given abundances of Bacteroides, Prevotella, and Ruminococcus. With a binary response, the predicted value $\hat{y}^*$ always lie in $[0, 1]$, so we assigned to a new data point the class label according to whether $\hat{y}^* > 0.5$. As before, we performed the study by randomly partitioning the 33 samples into training and test sets. Specifically, we set two-thirds of the observations from the European class and two-thirds of the observations from the other class as training samples, and the rest as test samples. We compared PAMIR with logistic regression. The test error rates, based on 100 random partitions, are summarized in Figure \ref{fig4}. We see that the error rate of logistic regression, which is higher than that of PAMIR, is about 33\%, which can be achieved by the classifier that assigns every observation to the European class. The results are similar as we vary the cutoff, see Table \ref{cutpoints}. The superior performance of PAMIR may be due to fact that inverse regression based methods can capture both linear and nonlinear patterns, while logistic regression is an inherently linear method.

\begin{table}[htb!]\caption{The average prediction error rate for logistic regression and PAMIR, over 100 random splits of 33 samples, are reported for a set of cutoffs.} \label{cutpoints} \vspace{-0.3cm}
\centering
 {\small\scriptsize
\renewcommand{\arraystretch}{1} 
\begin{tabular}{ccc}
\hline
Cutoff & Logistic regression & PAMIR \\
0.3  &  0.406 & 0.378 \\
0.4  &  0.374 & 0.300 \\
0.5  &  0.325 & 0.254 \\
0.6  &  0.341 & 0.270 \\
0.7  &  0.335 & 0.305 \\
\hline
\end{tabular} }
\end{table}

\section{Discussion}\label{sec_disc}

We developed a new method, PAMIR, for prediction analysis of microbiome sequencing data that accounts for the inherent properties of the data. An inverse regression model was proposed by reversing the trait of interest (i.e., the response) and bacterial counts (i.e., the covariates) in the regression. The underlying distribution for counts combines Aitchison's logistic normal distribution with the multinomial distribution \citep{billheimer2001statistical}. Model fitting was done via a Monte Carlo expectation-maximization algorithm, and predictions were made by exploiting the dimension-reduction structure in the model. 

In the E-step, the MH algorithm is used to compute the conditional expectation, for each observed value of the response. Hence, taking the tuning of the MH step size into account, parameter estimation can be very slow when the sample size is large. Fortunately, the fitting procedure enables parallel computing, and the computational time can be substantially reduced if multicore processors are available. We have implemented the proposed method in \textbf{R}, with parallel computing facilitated by the \textbf{snowfall} Package \citep{Snowfall2015}. Our limited experience suggests that the procedure works reasonably fast.  


Recent application of inverse regression in data mining applications include analysis of sentiment in text in social sciences \citep{taddy2010multinomial} and genome-wide test of associations in modern genetics \citep{song2015testing}. We anticipate that our inverse regression-based method will be useful for metagenomic studies. Also, the general framework could be modified to suit other domains, including genomics and proteomics.

\section{Appendix}

{\sc Proof of Proposition \ref{prop}.} Let $\bfXi_1 = \bfSigma^{-1/2} \bfGamma$ and let $\bfXi_2$ denote a complement of $\bfXi_1$ such that $(\bfXi_1, \bfXi_2)$ is a $(p - 1) \times (p - 1)$ orthogonal matrix. We have
$$\bfGamma^{\top} \bfSigma^{-1} \bfitW = \bfXi_1^{\top} \bfSigma^{-1/2} \bfitW$$
and
$$\bfSigma^{-1/2} \bfitW = \bfXi_1 \bfXi_1^{\top} \bfSigma^{-1/2} \bfitW + \bfXi_2 \bfXi_2^{\top} \bfSigma^{-1/2} \bfitW.$$
Hence
\begin{eqnarray*}
&& \{\bfitW \leq \bfitw \mid Y = y, \bfGamma^{\top} \bfSigma^{-1} \bfitW = \bfitu\} \\
&=& \{\bfXi_1 \bfXi_1^{\top} \bfSigma^{-1/2} \bfitW + \bfXi_2 \bfXi_2^{\top} \bfSigma^{-1/2} \bfitW \leq  \bfSigma^{-1/2} \bfitw \mid Y = y, \bfXi_1^{\top} \bfSigma^{-1/2} \bfitW = \bfitu\} \\
&=& \{\bfXi_2 \bfXi_2^{\top} \bfSigma^{-1/2} \bfitW \leq  \bfSigma^{-1/2} \bfitw - \bfXi_1 \bfitu \mid Y = y, \bfXi_1^{\top} \bfSigma^{-1/2} \bfitW = \bfitu\}.
\end{eqnarray*}
By (\ref{linear}),
\begin{equation*}
\bfXi_1^{\top} \bfSigma^{-1/2} \bfitW_y = \bfXi_1^{\top} \bfSigma^{-1/2} \bfmu + \bfbeta \bfith_y + \bfXi_1^{\top} \bfSigma^{-1/2} \bfxi
\end{equation*}
and
\begin{equation*}
\bfXi_2^{\top} \bfSigma^{-1/2} \bfitW_y =  \bfXi_2^{\top} \bfSigma^{-1/2} \bfmu + \bfXi_2^{\top} \bfSigma^{-1/2} \bfxi.
\end{equation*}
Since $\bfSigma^{-1/2} \bfxi$ is normally distributed with covariance matrix $\I_p$, $\bfXi_1^{\top} \bfSigma^{-1/2} \bfxi$ and $\bfXi_2^{\top} \bfSigma^{-1/2} \bfxi$ are independent. Consequently,
\begin{eqnarray*}
\operatorname{Pr}(\bfitW \leq \bfitw \mid Y = y, \bfGamma^{\top} \bfSigma^{-1} \bfitW = \bfitu) &=& \operatorname{Pr}(\bfXi_2 \bfXi_2^{\top} \bfSigma^{-1/2} \bfitW \leq  \bfSigma^{-1/2} \bfitw - \bfXi_1 \bfitu \mid Y = y) \\
&=& \operatorname{Pr}(\bfXi_2 \bfXi_2^{\top} \bfSigma^{-1/2} \bfitW_y \leq  \bfSigma^{-1/2} \bfitw - \bfXi_1 \bfitu) \\
&=& \operatorname{Pr}(\bfXi_2 \bfXi_2^{\top} \bfSigma^{-1/2} \bfitW \leq  \bfSigma^{-1/2} \bfitw - \bfXi_1 \bfitu) \\
&=& \operatorname{Pr}(\bfitW \leq \bfitw \mid \bfGamma^{\top} \bfSigma^{-1} \bfitW = \bfitu).
\end{eqnarray*}
Another way of saying this is that, given $\bfGamma^{\top} \bfSigma^{-1} \bfitW$, $Y$ is independent of $\bfitW$. Thus, the distribution of $Y \mid \bfitW$ is the same as the distribution of $Y \mid \bfGamma^{\top} \bfSigma^{-1} \bfitW$. The proof is complete.


{\sc Derivation of (\ref{update_gamma}) and (\ref{update_beta}).} It suffices to minimize
\begin{eqnarray*}
J(\bfGamma, \bfbeta) &=& \sum_{y} \sum_{b = 1}^{B}\{(\bfitw_{y}^{b} - \bar{\bfitw} - \bfGamma{\bfbeta} \bfith_y)^{\top} \bfSigma^{-1} (\bfitw_{y}^{b} - \bar{\bfitw} - \bfGamma {\bfbeta} \bfith_y)\}
\end{eqnarray*}
with respect to $(\bfGamma, \bfbeta)$. Note that
\begin{eqnarray*}
J(\bfGamma, \bfbeta) &=& c - 2 \sum_{y} \sum_{b = 1}^{B} (\bfitw_{y}^{b} - \bar{\bfitw})^{\top} \bfSigma^{-1} \bfGamma {\bfbeta} \bfith_y + \sum_{y} \sum_{b = 1}^{B} \bfith_y^{\top} {\bfbeta}^{\top} \bfGamma^{\top} \bfSigma^{-1} \bfGamma \bfbeta \bfith_y \\
&=& c - 2 B \sum_{y} (\bar{\bfitw}_{y} - \bar{\bfitw})^{\top} \bfSigma^{-1} \bfGamma {\bfbeta} \bfith_y + B \sum_{y} \bfith_y^{\top} {\bfbeta}^{\top} \bfbeta \bfith_y \\
&=& c - 2 B \sum_{y} {\operatorname{trace}}\{\bfSigma^{-1} \bfGamma {\bfbeta} \bfith_y (\bar{\bfitw}_{y} - \bar{\bfitw})^{\top}\} + B \sum_{y} {\operatorname{trace}}(\bfbeta \bfith_y \bfith_y^{\top} {\bfbeta}^{\top}) \\
&=& c - 2 B {\operatorname{trace}}\{\bfSigma^{-1} \bfGamma {\bfbeta} \H (\bar{\W} - \bar{\bfitw} \otimes {\bf 1}_n^{\top})^{\top}\} + B  {\operatorname{trace}}(\bfbeta \H\H^{\top} {\bfbeta}^{\top}).
\end{eqnarray*}
Here, $c$ is an irrelevant constant. Taking the derivative of $J(\bfGamma, \bfbeta)$ with respect to $\bfbeta$, and setting it equal to zero, we obtain
\begin{eqnarray*}
\H (\bar{\W} - \bar{\bfitw} \otimes {\bf 1}_n^{\top})^{\top} \bfSigma^{-1} \bfGamma = \H\H^{\top} \bfbeta^{\top},
\end{eqnarray*}
and hence
\begin{eqnarray*}
{\bfbeta} = \bfGamma^{\top}\bfSigma^{-1} (\bar{\W} - \bar{\bfitw} \otimes {\bf 1}_n^{\top}) \H^{\top} (\H\H^{\top})^{-1}.
\end{eqnarray*}
Plugging this into $J(\bfGamma, \bfbeta)$ and after some further manipulations, one needs to maximize
\begin{eqnarray*}
&& {\operatorname{trace}}\{\bfGamma^{\top}\bfSigma^{-1} (\bar{\W} - \bar{\bfitw} \otimes {\bf 1}_n^{\top}) \H^{\top} (\H\H^{\top})^{-1} \H (\bar{\W} - \bar{\bfitw} \otimes {\bf 1}_n^{\top})^{\top} \bfSigma^{-1}\bfGamma\} \\ &=& {\operatorname{trace}}\{\bfGamma^{\top}\bfSigma^{-1} \M \bfSigma^{-1}\bfGamma\}
\end{eqnarray*}
with respect to $\bfGamma$. Let $\bfitv_j(\bfSigma)$ denote the $j$th largest eigenvector of $\bfSigma^{-1/2} \M \bfSigma^{-1/2}$. Then the minimizer of $\bfGamma$ is
\begin{eqnarray*}
\tilde{\bfGamma} = \bfSigma^{1/2} \{\bfitv_1(\bfSigma), \ldots, \bfitv_d(\bfSigma)\},
\end{eqnarray*}
and the minimizer of $\bfbeta$ is
\begin{eqnarray*}
\tilde{\bfbeta} 
= \{\bfitv_1(\bfSigma), \ldots, \bfitv_d(\bfSigma)\}^{\top} \bfSigma^{-1/2} (\bar{\W} - \bar{\bfitw} \otimes {\bf 1}_n^{\top}) \H^{\top} (\H\H^{\top})^{-1}.
\end{eqnarray*}

\bibliographystyle{agsm}
\bibliography{bib}

\end{document}